\documentclass[10pt,twocolumn,letterpaper]{article}
\usepackage{geometry}
\usepackage{times}
\usepackage{epsfig}
\usepackage{graphicx}
\usepackage{amsmath}
\usepackage{amssymb}

\usepackage{hyperref}
\usepackage{array} 
\usepackage{booktabs}
\usepackage{siunitx} 

\begin{document}

\title{\centering Validation of Practicality for CSI Sensing Utilizing Machine Learning}

\author{Tomoya, Tanaka\\
Georgia Institute of Technology\\
{\tt\small ttanaka9@gatech.edu}\\
SoftBank Corp.\\
{\tt\small tomoya.tanaka02@g.softbank.co.jp}
\and
Ayumu, Yabuki\\
SoftBank Corp.\\
{\tt\small ayumu.yabuki@g.softbank.co.jp}
\and
Mizuki, Funakoshi\\
SoftBank Corp.\\
{\tt\small mizuki.funakoshi@g.softbank.co.jp}
\and
Ryo, Yonemoto\\
SoftBank Corp.\\
{\tt\small ryo.yonemoto@g.softbank.co.jp}
}

\maketitle

\begin{abstract}
   In this study, we leveraged Channel State Information (CSI), commonly utilized in WLAN communication, as training data to develop and evaluate five distinct machine learning models for recognizing human postures: "standing," "sitting," and "lying down." The models we employed were: (i) Linear Discriminant Analysis, (ii) Naive Bayes-Support Vector Machine, (iii) Kernel-Support Vector Machine, (iv) Random Forest, and (v) Deep Learning. We systematically analyzed how the accuracy of these models varied with different amounts of training data. Additionally, to assess their spatial generalization capabilities, we evaluated the models' performance in a setting distinct from the one used for data collection. The experimental findings indicated that while two models— (ii) Naive Bayes-Support Vector Machine and (v) Deep Learning—achieved 85\% or more accuracy in the original setting, their accuracy dropped to approximately 30\% when applied in a different environment. These results underscore that although CSI-based machine learning models can attain high accuracy within a consistent spatial structure, their performance diminishes considerably with changes in spatial conditions, highlighting a significant challenge in their generalization capabilities.
\end{abstract}

\section{Introduction}
In recent years, the field of human recognition has seen significant advancements, particularly with the application of deep learning models to camera-based systems, enabling highly accurate object identification. These technologies have already been successfully implemented across various domains, including security, healthcare, and smart environments. However, in indoor environments such as households, where privacy concerns are paramount, the use of cameras can be problematic. As a result, there has been a growing interest in alternative methods that leverage sensors for human recognition. Among these methods, one approach that has gained considerable attention is the use of Channel State Information (CSI) obtained from WLAN devices for sensing applications.

CSI, which represents the state of the propagation path between a transmitter and receiver in WLAN communication, provides a wealth of multidimensional data regarding amplitude and phase displacement across multiple antennas. This data is particularly valuable because WLAN signals interact with the human body, being either blocked or attenuated as they pass through, which allows CSI to capture information that can be used for object and human recognition. The ability to use existing WLAN infrastructure for this purpose presents a cost-effective and non-intrusive solution, making it an attractive area of research[1].

Several studies have explored methods for human recognition using CSI[2]-[50]. For instance, one study proposed a method that restricts the propagation path of WLAN signals and analyzes changes in CSI when a person crosses this path, demonstrating a novel approach to human recognition[51]. Additionally, another study successfully developed a deep learning encoder-decoder model trained on CSI data, which enabled the generation of Dense Pose representations of the human body[52]. These studies highlight the potential of CSI for human recognition and its advantages in scenarios where privacy is a concern.

Despite these promising developments, the practical application of CSI sensing remains significantly behind that of camera-based systems. While research outcomes have been impressive, with several papers indicating CSI's capability in various settings, the transition to real-world applications has been slow. One of the key challenges identified in several relevant papers in the field is the potential decrease in recognition accuracy when the spatial structure changes[53][54]. If this issue proves significant, it could be a major barrier to the widespread adoption of CSI sensing technologies.

To quantitatively examine this issue, we undertook a comprehensive and quantitative evaluation, constructing five different machine learning models to identify human postures such as "standing," "sitting," and "lying down." Our study specifically aimed to assess how changes in spatial structure impact the recognition accuracy of each model. Understanding the extent to which spatial variations affect CSI sensing accuracy is crucial for the future practical application of sensor-based object detection systems. Such insights will be vital in overcoming the current limitations and moving towards the broader implementation of sensor-based recognition technologies.

\section{Approach}
\subsection{Experimental Setup}
In this experiment, we used a computer connected to three antennas that receive signals from a WLAN router transmitting signals. The frequency band used was 2.4GHz, and the WLAN signals transmitted from the three antennas connected to the router were received as Channel State Information (CSI) data by the three receiving antennas. This CSI data was saved in binary DAT files on the computer connected to the receiving antennas. The saved DAT files were subsequently converted into 5-dimensional NUMPY files and used as training data for the machine learning models.

The converted NUMPY files have a shape of (5, 30, 3, 3, 2). In this structure, the first dimension represents the time slots of data acquisition (5 instances), the second dimension represents the subcarrier index, the third dimension represents the transmitting antenna index, the fourth dimension represents the receiving antenna index, and the fifth dimension contains the amplitude and phase data. This data structure allows for capturing the complex physical environmental changes during WLAN signal propagation in a multidimensional format, providing a rich set of features to enhance the performance of the machine learning models.

In the experiment, a single participant assumed the postures of "standing," "sitting," and "lying down" between the router and the receiving antennas, and the signal blocking and attenuation patterns associated with each posture were recorded(Figure 1). 

\begin{figure}[htbp]
\centering
\fbox{ 
\includegraphics[width=0.45\textwidth]{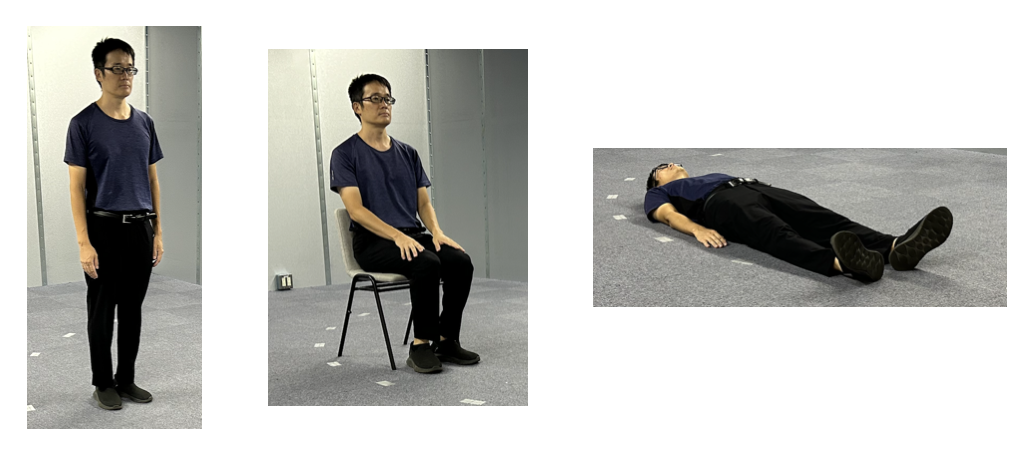} 
}
\caption{Postures: Stand, Sit, Lying Down}
\label{fig:postures: sit, stand, lying down}
\end{figure}

These patterns were extracted as features and used as training data for the machine learning models. This approach enabled the precise capture of the effects that different postures have on WLAN signals, aiming for high-precision posture recognition based on these effects.

A conceptual diagram of the test configuration is presented in Figure 2 below.

\begin{figure}[htbp]
\centering
\fbox{ 
\includegraphics[width=0.45\textwidth]{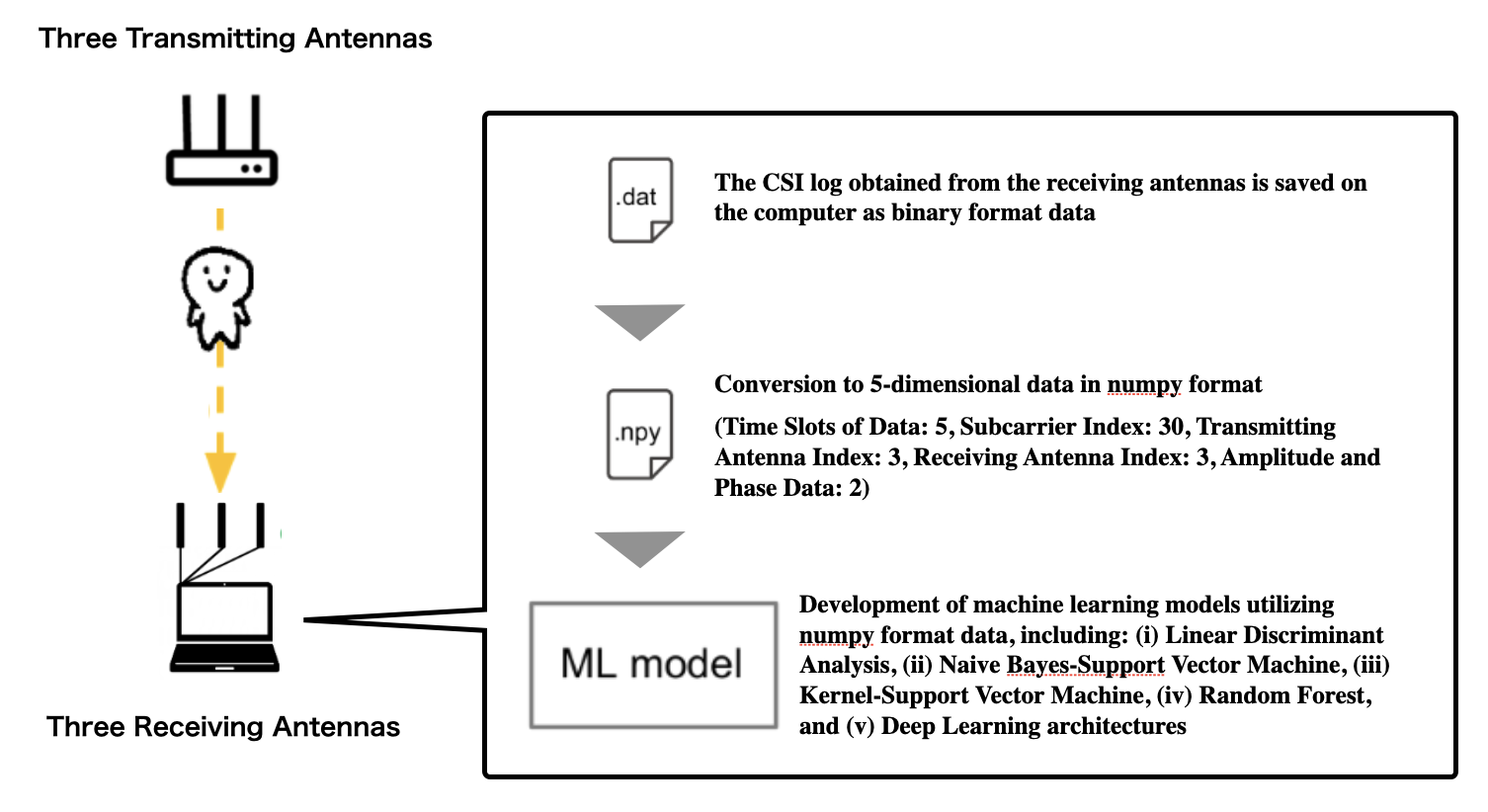} 
}
\caption{Conceptual Diagram of Test Configuration}
\label{fig:postures: sit, stand, lying down}
\end{figure}

\subsection{Machine Learning Models}
For this experiment, we constructed the following five machine learning models.

\subsubsection{Linear Discriminant Analysis, LDA}
Linear Discriminant Analysis (LDA) is a statistical method widely applied in multi-class classification problems. The primary objective of LDA is to identify the optimal linear boundary that separates data points belonging to different classes. This technique aims to improve classification performance by simultaneously maximizing between-class variance and minimizing within-class variance.
LDA classifies data using the following linear discriminant function y(x):
\\
\begin{equation}
y(\mathbf{x}) = \mathbf{w}^T \mathbf{x} + w_0
\label{eq:lda_function}
\end{equation}
\\
where x is the input feature vector, w is the weight vector, and $w_0$ is the bias term. In a two-class problem, the optimal weight vector w is calculated using the following equation:
\\
\begin{equation}
\mathbf{w} = \mathbf{S}^{-1}(\boldsymbol{\mu}_1 - \boldsymbol{\mu}_2)
\label{eq:lda_weight}
\end{equation}
\\
where S is the sum of the between-class covariance matrices, and $\mu_1$ and $\mu_2$ represent the mean vectors of each class.
LDA is particularly effective when features follow a normal distribution, and due to its computational efficiency and ease of interpretation, it has been widely adopted in various application fields. In this study, we constructed a posture classification model for three distinct postures using LDA.

\subsubsection{Naive Bayes Support Vector Machine, NB-SVM}
Naive Bayes Support Vector Machine (NB-SVM) is an innovative machine learning algorithm that combines the strengths of Naive Bayes classifiers and Support Vector Machines (SVM). This method demonstrates excellent performance, particularly with high-dimensional data, especially in text classification tasks.
The processing flow of NB-SVM is as follows:
\begin{enumerate}
    \item Feature preprocessing using Naive Bayes: Extraction of probabilistic information for each feature
    \item Use of extracted probability information as input for SVM
    \item Final class classification by SVM
\end{enumerate}

Naive Bayes is a probabilistic classification algorithm based on Bayes' theorem, which calculates the probability of data belonging to class Ck using the following equation:
\\
\begin{equation}
P(C_k|x) = \frac{P(x|C_k)P(C_k)}{P(x)}
\label{eq:naive_bayes}
\end{equation}
\\
Here, $P(C_k|x)$ is the posterior probability that feature vector $x$ belongs to class $C_k$, $P(x|C_k)$ is the likelihood of $x$ given class $C_k$, $P(C_k)$ is the prior probability of $C_k$, and $P(x)$ is the evidence. The characteristic of Naive Bayes is that it assumes all features are mutually independent and calculates the class membership probability using the product of probabilities for each feature. This method is computationally efficient and particularly effective for high-dimensional data.
On the other hand, Support Vector Machine (SVM) is an algorithm that finds the optimal hyperplane to separate different classes. SVM solves the following optimization problem to maximize the margin (the distance between the hyperplane and the closest data points):
\\
\begin{equation}
\text{maximize } \frac{2}{|w|}
\label{eq:svm_optimization}
\end{equation}
\\
Here, $w$ is the normal vector of the hyperplane. The optimal hyperplane is represented by the following equation:
\\
\begin{equation}
w \cdot x + b = 0
\label{eq:hyperplane}
\end{equation}
\\
NB-SVM aims to achieve better performance than conventional SVM by combining the efficient feature processing capability of Naive Bayes with the classification power of SVM. In this study, we apply NB-SVM to a posture detection model, aiming for high-accuracy posture recognition.

\subsubsection{Support Vector Machine with Kernel Trick}
The kernel trick is a sophisticated technique widely utilized in Support Vector Machines (SVM), known for its efficient resolution of non-linear separation problems. The essence of this method lies in implicitly mapping data that is difficult to separate linearly in the original space into a higher-dimensional feature space. This enables the treatment of complex classification tasks as linear separation problems while significantly reducing computational complexity.

The core of the kernel trick lies in the use of the kernel function $K(x_i, x_j)$. This function allows for direct computation of the inner product between data points $x_i$ and $x_j$ in a high-dimensional space. The general form of a kernel function is expressed as follows:
\\
\begin{equation}
K(x_i, x_j) = \phi(x_i) \cdot \phi(x_j)
\label{eq:kernel_function}
\end{equation}
\\
Here, $\phi(x)$ represents the mapping from the input space to the high-dimensional feature space. The innovation of the kernel trick lies in its ability to evaluate $K(x_i, x_j)$ directly without explicitly computing $\phi(x)$. This achieves non-linear separation while dramatically reducing computational costs.

Representative kernel functions include the following:
\begin{enumerate}
    \item Linear kernel:
    \vspace{5pt}
    \begin{equation}
    K(x_i, x_j) = x_i \cdot x_j
    \label{eq:linear_kernel}
    \vspace{5pt}
    \end{equation}
    
    \item Polynomial kernel:
     \vspace{5pt}
    \begin{equation}
    K(x_i, x_j) = (x_i \cdot x_j + c)^d
    \label{eq:polynomial_kernel}
    \vspace{5pt}
    \end{equation}
    
    \item Gaussian (RBF) kernel:
    \vspace{5pt}
    \begin{equation}
    K(x_i, x_j) = \exp(-\gamma \|x_i - x_j\|^2)
    \label{eq:gaussian_kernel}
    \vspace{5pt}
    \end{equation}
\end{enumerate}

The selection of these kernel functions is made based on the characteristics of the problem and the required degree of non-linearity.

In this study, we aim to improve SVM performance in posture recognition by selecting the optimal kernel function and adjusting parameters to realize a posture recognition model with higher accuracy than linear SVM. By applying the kernel trick, we expect to recognize complex patterns that conventional linear SVMs could not capture, leading to a significant improvement in classification accuracy.

\subsubsection{Random Forest}
Random Forest is a powerful ensemble learning algorithm that integrates multiple decision trees to make predictions. The fundamental component of this method, the Decision Tree, is a hierarchical classification and regression model excellent at capturing non-linear patterns. Each decision tree recursively partitions the feature space and classifies data points based on conditions.
In this study, we adopted decision trees based on Gini Impurity. Gini Impurity is a metric that quantifies the degree of class mixing at a node and is defined by the following equation:
\\
\begin{equation}
Gini(t) = 1 - \sum_{i=1}^{c} p(i|t)^2
\label{eq:gini_impurity}
\end{equation}
\\
Here, $Gini(t)$ represents the Gini impurity at node $t$, $c$ is the number of classes, and $p(i|t)$ is the proportion of data belonging to class $i$ at node $t$. At each branching point, the feature and threshold that maximize the decrease in Gini impurity ($\Delta$Gini) are selected:
\\
\begin{equation}
\Delta Gini = Gini(parent) - \sum_{j \in {left,right}} \frac{N_j}{N} Gini(j)
\label{eq:gini_gain}
\end{equation}
\\
$N$ denotes the number of samples in the parent node, and $N_j$ represents the number of samples in child node $j$.
Random Forest generates multiple decision trees based on this Gini impurity and aggregates their predictions, mitigating the overfitting tendency of a single decision tree and improving the model's generalization performance. Specifically, bootstrap sampling and random feature selection are applied during the learning of each decision tree to construct a diverse set of decision trees. The final prediction is determined by majority voting (for classification problems) or averaging (for regression problems) of the outputs from all decision trees.
In this research, we aim to apply this method to the posture recognition task to construct a high-accuracy and robust classification model. The Random Forest model, which utilizes bagging through bootstrap sampling, is expected to contribute to effective identification of diverse posture patterns by preventing overfitting and enhancing resilience against noise. This approach allows the model to capture complex interactions between features while maintaining robust performance.

\subsubsection{Deep Learning : Convolutional Neurarl Network(CNN) }

Convolutional Neural Network (CNN) is an advanced architecture specialized for image recognition and processing tasks within the deep learning paradigm. The excellence of CNN stems from its hierarchical structure, possessing the ability to effectively extract and learn spatial features of input data through the alternating arrangement of convolutional and pooling layers.
The core of CNN lies in the convolution operation, which generates feature maps through convolution operations between input data and learnable filters (kernels). This process is expressed by the following equation:
\\
\begin{equation}
F(x,y) = (I * K)(x,y) = \sum_{i=-a}^{a} \sum_{j=-b}^{b} I(x-i, y-j)K(i,j) + b
\label{eq:convolution}
\end{equation}
\\
Here, \( F(x, y) \) represents the output feature map, \( I(x, y) \) is the input data, \( K(i, j) \) is the convolution kernel, and \( b \) is the bias term. This operation extracts local features, which are then transformed into higher-order representations in subsequent layers. These feature maps are typically passed through pooling layers, where downsampling operations, such as max pooling, reduce the spatial dimensions of the feature maps, preserving the most important information while decreasing computational complexity. Eventually, these condensed feature maps are fed into fully connected layers, where they are flattened and used in the final classification process, enabling the model to make accurate predictions based on the learned features.

In this study, we optimized the input shape of the CNN model considering the characteristics of Channel State Information (CSI) data. Specifically, we first reduced the conventional 5-dimensional data structure (5,30,3,3,2) to a 4-dimensional structure (5,30,3,3) by using only the amplitude information and discarding the phase data. This transformation aims to avoid instability associated with simultaneous learning of data with different characteristics (amplitude and phase) and improve model convergence by focusing solely on amplitude data. Subsequently, we transformed this into a 3-dimensional structure (5,30,9) by integrating the transmit and receive antenna combinations. Furthermore, through batch processing and dimension rearrangement, we set the final input shape to (batch size, 9, 30, 5). 

The architecture of the constructed CNN model is shown in Figure 3.

\begin{figure}[htbp]
\centering
\fbox{ 
\includegraphics[width=0.4113\textwidth]{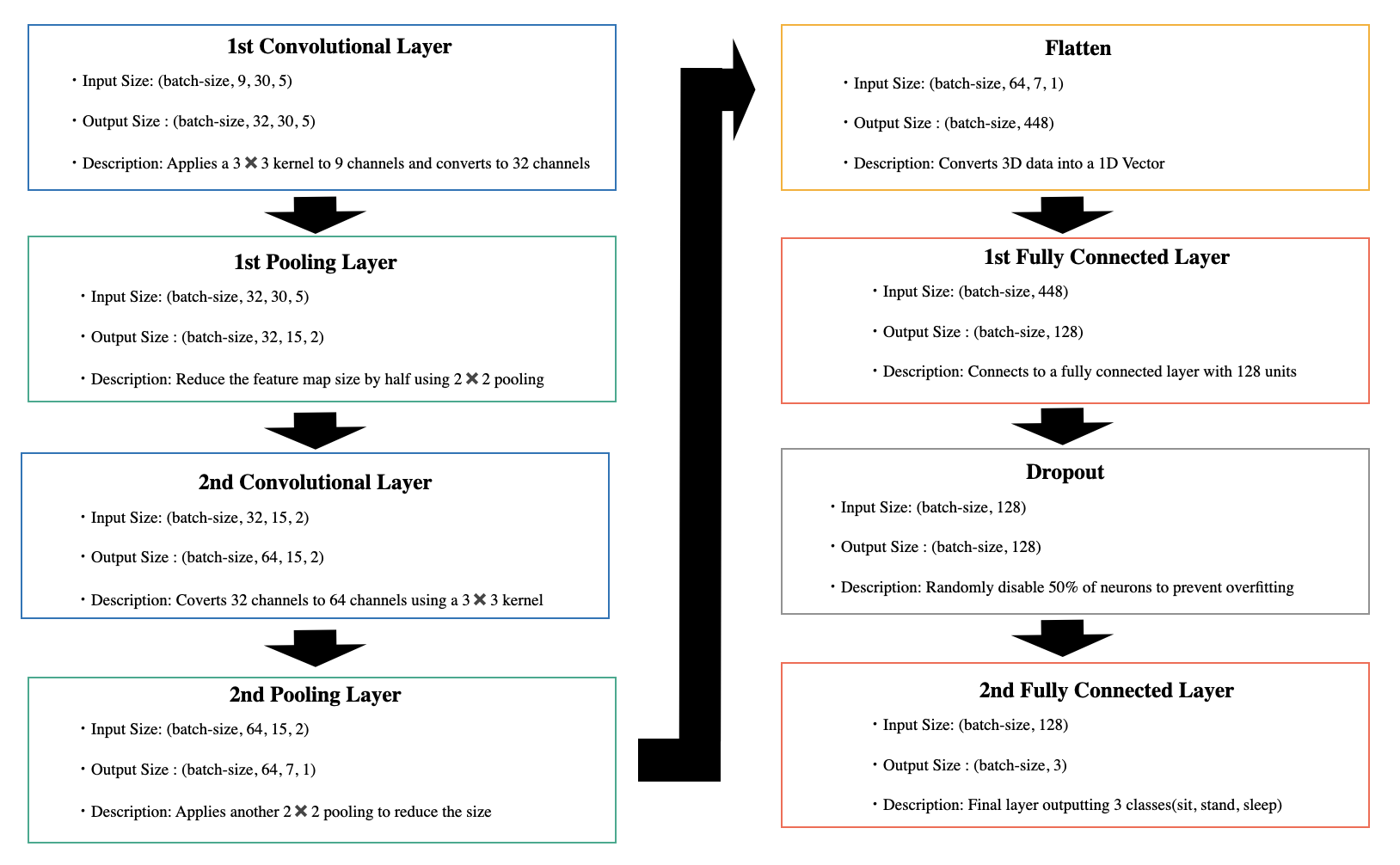} 
}
\caption{Architecture of the CNN model}
\label{fig:postures: sit, stand, lying down}
\end{figure}

This model configuration is expected to effectively capture the temporal and spatial features inherent in CSI data, achieving high-accuracy posture recognition. Moreover, this approach is believed to enable the acquisition of more robust and versatile feature representations compared to conventional CSI-based posture recognition methods.

\subsection{Learning and Validation Protocol for Machine Learning Models}

In this study, we adopted the following protocol to conduct rigorous and systematic learning and validation of machine learning models.

\subsubsection{Data Collection Environments}

The experiments were conducted in two different environments:

\begin{enumerate}
    \item Environment A: Indoor space measuring 5m in width, 3m in length, and 2.5m in height
    \item Environment B: Indoor space measuring 6.6m in width, 4.7m in length, and 2.6m in height
\end{enumerate}

Although these rooms differ in size and shape, the distance and height of the transmitting router and receiving antenna were kept consistent in both rooms. This consistency was crucial for accurately assessing the impact of different spatial configurations on the identification accuracy of the machine learning models.

\begin{figure}[htbp]
\centering
\fbox{ 
\includegraphics[width=0.45\textwidth]{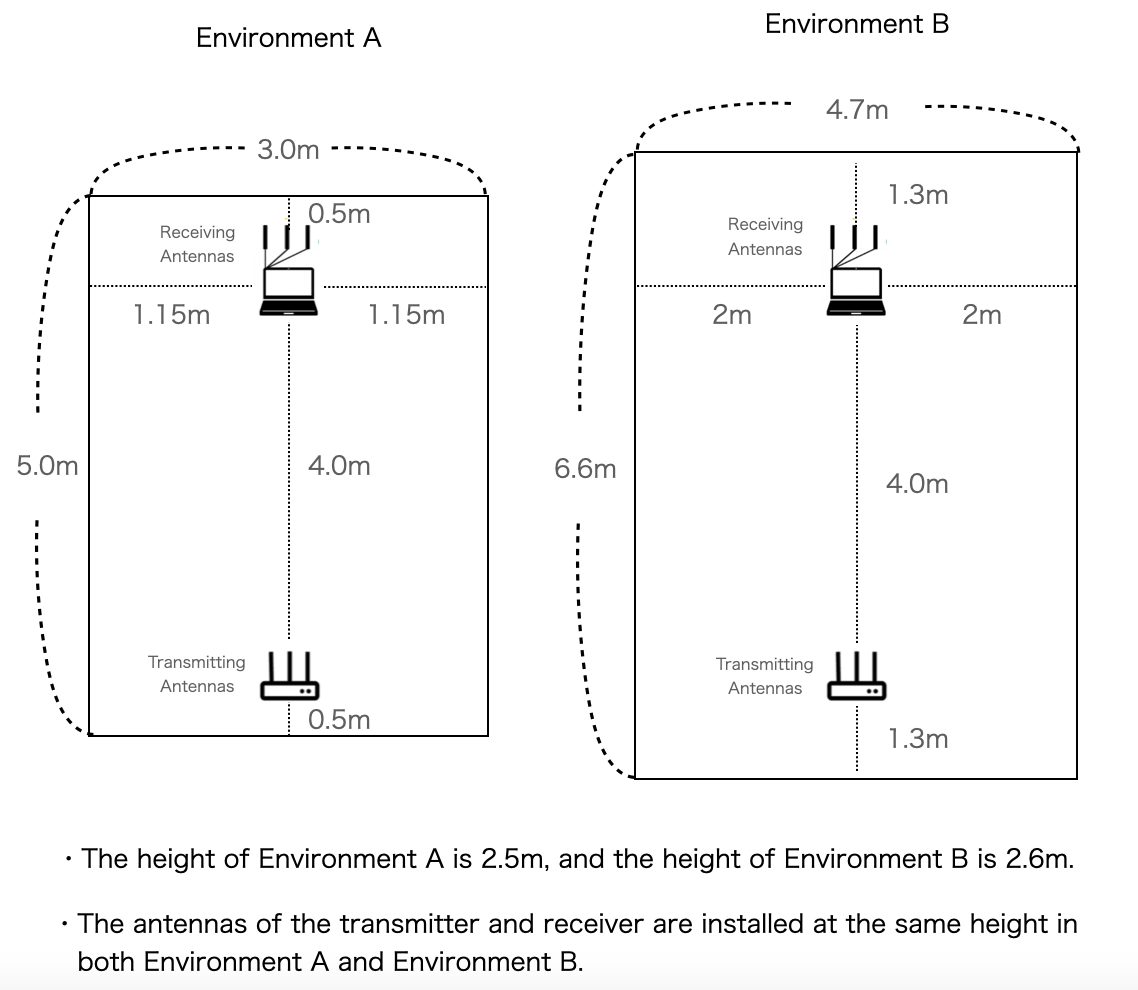} 
}
\caption{Layouts of Environment A and B}
\label{fig:postures: sit, stand, lying down}
\end{figure}

\subsubsection{Data Collection Process}

In Environment A, we collected 2,000 samples of CSI data for each of the three basic postures ("standing", "sitting", "lying down"), totaling 6,000 samples. During data collection, we considered the diversity of subjects (age, gender, body type) to ensure comprehensiveness and representativeness of the CSI data.

\subsubsection{Dataset Splitting and Learning Set Configuration}

We divided the collected dataset as follows for learning and validation:

\begin{enumerate}
    \item Validation dataset: 200 samples (600 samples in total) were randomly extracted from each posture category and reserved for validation.
    \item Learning dataset: Based on the remaining 1,800 samples per category (5,400 samples in total), we constructed six types of learning datasets as follows:
    \begin{itemize}
        \item Set A: 300 samples per posture (900 samples in total)
        \item Set B: 600 samples per posture (1,800 samples in total)
        \item Set C: 900 samples per posture (2,700 samples in total)
        \item Set D: 1,200 samples per posture (3,600 samples in total)
        \item Set E: 1,500 samples per posture (4,500 samples in total)
        \item Set F: 1,800 samples per posture (5,400 samples in total)
    \end{itemize}
\end{enumerate}

This gradual dataset configuration allowed us to quantitatively evaluate the impact of learning data volume on model performance.

\subsubsection{Model Learning and Performance Evaluation Process}

We applied the following process to each machine learning model:

\begin{enumerate}
    \item We trained the models using each learning dataset from Set A to F.
    \item We evaluated the identification accuracy of each trained model using the validation dataset (600 samples) extracted from Environment A.
\end{enumerate}

\subsubsection{Generalization Performance Evaluation Across Environments}

To evaluate the generalization performance of the models, we conducted the following process:

\begin{enumerate}
    \item In Environment B, we newly collected 100 samples of CSI data for each of the three basic postures, totaling 300 samples.
    \item We verified the identification accuracy using the 300 samples collected in Environment B for each model trained with Set F from Environment A (1,800 samples per posture, 5,400 samples in total).
\end{enumerate}

This evaluation process allowed us to verify the performance of the models in an environment with different spatial characteristics from the learning environment, confirming their effectiveness from a practical perspective.

\subsubsection{Performance Metrics}

For the evaluation of model performance, we employed the following metrics:

\begin{itemize}
    \item Overall Accuracy
    \item Precision, Recall, and F1 Score (for each posture category)
\end{itemize}

By utilizing these multifaceted evaluation metrics, we were able to comprehensively analyze the performance of the models and clarify the strengths and weaknesses in identifying each posture.

\section{Experimental Results}
\subsection{Evaluation in Environment A}
The experiments conducted on the models, including (i) Linear Discriminant Analysis, (ii) Naive Bayes-Support Vector Machine, (iii) Kernel-Support Vector Machine, (iv) Random Forest, and (v) Deep Learning, revealed that very high identification accuracy was achieved when the accuracy was evaluated under the same conditions as Environment A, where the training data was collected. Specifically, as shown in the Accuracy section of Figure 5, the identification accuracy of models built using Set F (a total of 5400 data points) was 85\% or more for both the (ii) Naive Bayes-Support Vector Machine model and the (v) Deep Learning model, confirming the superior performance of these models in Environment A. As observed from the F1 scores, all models trained on 5400 data points demonstrated relatively higher accuracy in identifying Sleep, while the accuracy for identifying Stand was lower.

This result suggests that, due to the high-dimensional and dense nature of CSI data, it is possible to build highly accurate models even with a relatively small amount of data. This can be considered a significant advantage of CSI sensing technology.

\subsection{Evaluation in Environment B}
On the other hand, when accuracy was evaluated using validation data from a different environment, Environment B, the identification accuracy of all models significantly decreased to approximately 30\%, as shown in the Accuracy section of Figure 6. This result indicates that machine learning models using CSI have very low spatial generalization capability, and even slight changes in the environment (such as changes in furniture arrangement or human position) can lead to a substantial decrease in identification accuracy.

\subsection{Discussion of Results}
These results highlight a major challenge in the practical application of CSI sensing. In particular, it is essential to develop new approaches to improve spatial generalization capabilities to enable high-accuracy identification in different environments.

\begin{figure}[htbp]
\centering
\fbox{ 
\includegraphics[width=0.45\textwidth,height=10.2cm]{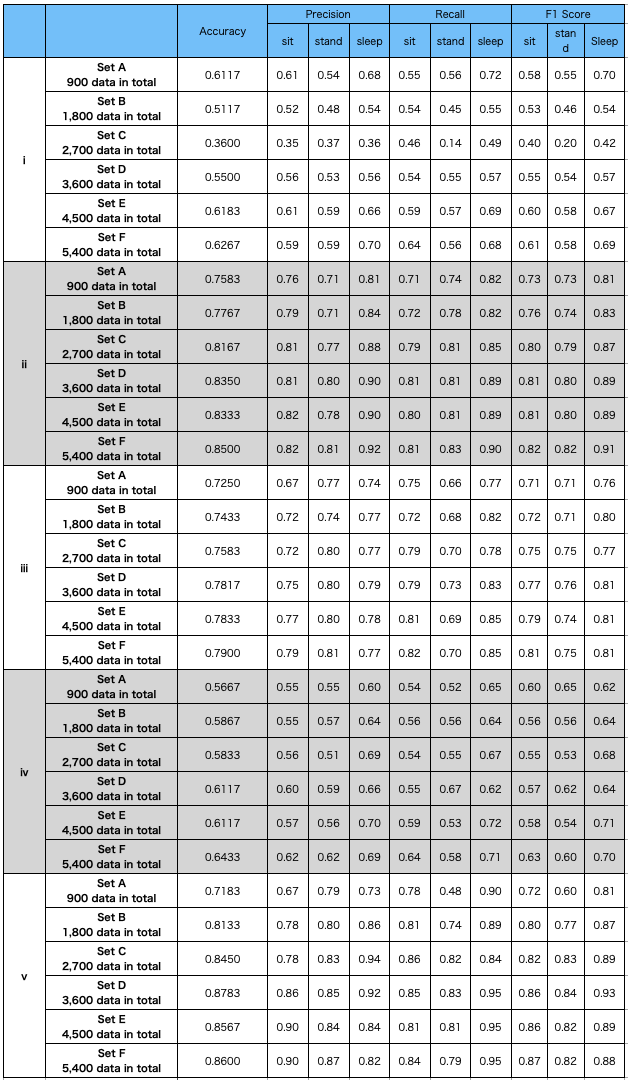} 
}
\caption{Model Accuracy in Environment A}
\label{fig:postures: sit, stand, lying down}
\end{figure}

\begin{figure}[htbp]
\centering
\fbox{ 
\includegraphics[width=0.45\textwidth]{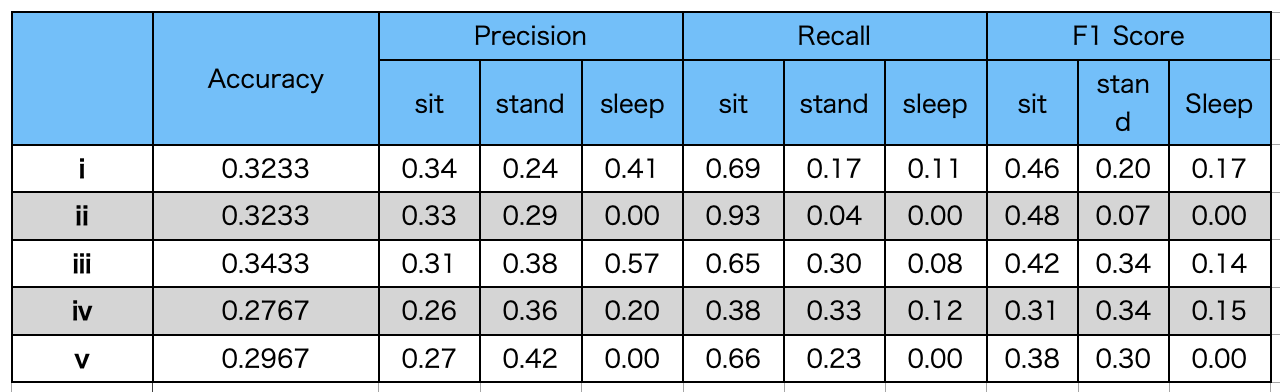} 
}
\caption{Model Accuracy in Environment B}
\label{fig:postures: sit, stand, lying down}
\end{figure}

\section{Conclusion}
In this paper, we conducted a detailed investigation into how the performance of human posture recognition models based on WLAN CSI information is affected by changes in spatial structures. Our experimental results revealed that the accuracy of these models significantly decreases under different spatial configurations, highlighting a major challenge in the application of WLAN CSI-based sensing technology.

To address this phenomenon and move towards the practical application of radio wave sensing technology, it is essential to undertake the following multifaceted approaches. First, (i) the development of methods for dynamically correcting signal characteristics in each environment is required. This would enable the models to adapt to varying spatial characteristics and maintain accuracy. Next, (ii) incorporating samples from diverse environments into the training dataset is expected to enhance the generalization ability of the models, allowing them to achieve high recognition accuracy across various settings. Additionally, (iii) the development of CSI sensing transmission and reception equipment with enhanced WLAN signal directivity is an important step. This would help reduce noise caused by signal scattering and reflection, thereby improving accuracy. Finally, (iv) considering the use of RADAR beamforming technology, which offers higher directivity than the CSI-based method, as an alternative is also worth exploring. The use of RADAR could allow for more precise environmental sensing and signal analysis, potentially leading to improved posture recognition accuracy.

This research underscores not only the current limitations but also the significant potential of radio wave sensing technology. It brings to light important areas that need further exploration and refinement, thereby suggesting new and promising directions for future research and development. By identifying these critical aspects, the study sets the stage for advancements that could bridge the existing gaps and propel the technology forward. Building on the insights gained so far, it is evident that further technological innovation, coupled with the adoption of multifaceted and interdisciplinary approaches, will be essential. These efforts are not merely incremental but foundational to enhancing the robustness, reliability, and accuracy of radio wave sensing systems.

These advancements will pave the way for the broader practical application of radio wave sensing technology, enabling its use across various fields in the future. Areas such as healthcare, smart homes, security systems, and environmental monitoring will benefit from the non-intrusive and flexible nature of radio wave-based solutions. Overcoming the challenges identified in this research is essential for building effective posture recognition systems. Furthermore, continuous progress through sustained research and development is necessary to fully unlock the potential of radio wave sensing and ensure its usefulness in real-world applications. Efforts in this field will help make the technology more practical and widely integrated into our daily lives.

\end{document}